\documentclass[12pt,a4paper]{article}

\usepackage[cp1251]{inputenc}
\usepackage[T1]{fontenc}
\usepackage[table]{xcolor}
\usepackage{url}
\usepackage{color}
\usepackage{amssymb}
\usepackage{amsmath}
\usepackage{amsfonts}

\usepackage{array,longtable}

\newcommand{\R}{\mathbb{R}}
\newcommand{\cC}{\mathbb{C}}

\newcommand{\Gr}{Gr\"obner }

\evensidemargin\oddsidemargin
\advance\textheight-\headheight
\advance\textheight-\headsep
\advance\textheight-\topskip
\title{On the ring of local unitary invariants for mixed $X-$states of two qubits}

\date{}

\author{
Vladimir P.Gerdt,\ \ Arsen M. Khvedelidze and Yuri G. Palii
}

\begin{document}

\maketitle

\begin{abstract}
Entangling properties of a mixed 2-qubit system can be described by the local homogeneous unitary invariant polynomials in elements of the density  matrix. The structure of the corresponding invariant polynomial ring for the special subclass of states, the so-called mixed $X-$states, is established. It is shown that for the $X-$states there is an injective ring homomorphism of the quotient ring of $SU(2)\times SU(2)$ invariant polynomials modulo its syzygy ideal and the $SO(2)\times SO(2)-$invariant ring freely generated by five homogeneous polynomials of degrees 1,1,1,2,2.
\end{abstract}

\maketitle
\tableofcontents
\section{Introduction}

\noindent{ $\bullet$\, \bf  Motivation $\bullet$ }In this paper, we consider a bipartite quantum system composed of two qubits, whose state
space, $\mathfrak{P}_X, $ is a special 7-dimensional family of the so-called
$X-$states \cite{YuEberly2007}. Our interest to this subspace of a generic two qubit space
 $\mathfrak{P}$ is due to fact that many well-known states, e.g. the
Bell states \cite{NielsenChuang}, Werner states \cite{Werner},
isotropic states \cite{HorodeckiHorodecki99}  and maximally entangled mixed states
\cite{IshizakaHiroshima2000,VerstraeteAudenaertBieMoor2001} are particular subsets of the $X-$states.
Since their introduction in \cite{YuEberly2007} many  interesting properties of  $X-$states have been established. Particularly,  it was shown that for a fixed set of eigenvalues the states of maximal concurrence, negativity or relative entropy of entanglement are the $X-$states.~\footnote{ For detailed review of the $X-$states and their applications we refer to the recent article \cite{MendoncaMarchiolliGaletti2014}.}

\noindent{ $\bullet$\, \bf  Contents  $\bullet$ }Here we pose the question about the algebraic structure of the local unitary polynomial invariants algebra corresponding to the $X-$states. More precisely, the fate of  generic  $SU(2)\times SU(2)$\--invariant polynomial ring of 2-qubits
\cite{Quesne}\---\cite{GerdtKhvedelidzePalii2009}
under the restriction  of the total  2-qubits  state space $\mathfrak{P} $ to its subspace $\mathfrak{P}_X$ will be discussed.
The quotient structure of the ring obtained as a result of restriction will be determined.
Furthermore, we establish an injective homo\-mor\-phism between this ring
and the invariant ring ${\R}[\mathfrak{P}_X]^{SO(2)\times SO(2)}\,, $ of
local unitary invariant polynomials for 2-qubit  $X-$states. In doing so, we  show that the latter ring is \textit{freely} generated by  five  homogeneous  invariants of degrees 1,1,1,2,2.


\section{Framework and settings}

In this section the collection of main algebraic  structures  associated
with a finite dimensional quantum system is given.

\subsection{General algebraic settings and conventions}

Hereafter, we use the standard notation, $\R[x_1, \ldots x_n ]\,,$ for the ring of polynomials in $n$ variables $x_1, \ldots x_n $  with coefficients in $\R\,.$
Given a polynomial set
\begin{equation}
F:=\{\,f_1,\dots, f_m\,\} \in \R[x_1,\dots, x_n]\,, \label{PolSet}
\end{equation}
generating the subring
\begin{equation}\label{subring}
\R[F]:=\R[f_1,\dots, f_m ] \subset \R[x_1,\ldots,  x_n ],
\end{equation}
we shall consider the polynomial ring $\R[y_1,\ldots,y_m]$ associated with $\R[F]$ where $y_1,\ldots,y_m$ are variables (indeterminates).

Note, that $\R[F]$ differs from the \textit{ideal}
$I_F =\langle F \rangle\subseteq \R[x_1,\ldots,  x_n ]\,,$
generated by $F$
\begin{equation}
I_F=\left\{ \sum_{i=1}^m h_i f_i \,\mid \,  h_1, \dots, h_m \in \mathbb{R}[x_1,\ldots, x_n ]\right\}\,. \label{ideal}
\end{equation}
The polynomial set $F$ defines the real affine variety $V\subset \R^m$. The radical ideal $I(V):=\sqrt{I_F}$ of $I_F$, i.e. the ideal such that $f\in \sqrt{I_F}$ iff $f^m\in I_F$ for some positive integer $m$, yields a {\em coordinate ring} of $V$ as the quotient ring
\begin{equation}
     \mathbb{R}[x_1,\ldots,  x_n ]/I(V)\,. \label{CoorRing}
\end{equation}
A nonzero polynomial $g\in \R[y_1,\ldots,y_m]$ such that $g(f_1,\ldots,f_m)=0$ in $\R[x_1, \ldots,  x_n ]$ is called a {\it syzygy} or a {\it nontrivial algebraic relation} among $f_1,\ldots,f_m$. The set of all syzygies forms the {\em syzygy ideal}
\[
I_F:=\{\,s\in \R[y_1,\ldots,y_m]\ \mid\ s=0\ \text{in}\ \R[x_1,\ldots,x_n]\,\}\,.
\]
In doing so, the following ring isomorphism holds (cf.~\cite{CLO'07}, Ch.7, Prop.2)
\begin{equation}
     \R[F] \cong \R[y_1,\ldots,y_m]/I_F\,. \label{isomorphism}
\end{equation}
Given an ideal $I_F$ in~\eqref{ideal}, a subset $\mathfrak{X}\subseteq \{\,x_1,\ldots,x_n\,\}$ of indeterminates is called
{\em independent modulo $I_F$} if $I_F \cap \R[\mathfrak{X}] = \{\,\}$. Otherwise, $\mathfrak{X}$ is called {\em dependent modulo
$I_F$}. The {\em affine dimension} of $I_F$, denoted by $\dim(I_F)$, is defined to be
the cardinality of a largest subset independent modulo $I_F$. If $I_F = \R[x_1,\ldots,x_n]=\langle 1\rangle$, then the affine dimension of $I_F$ is defined to be $-1$.

The ring of elements in $\mathbb{R}[x_1,\ldots, x_n ]$ which are invariant under the action of a group $G$ on $\{\,x_1,\ldots,x_n\,\}$ will be denoted by $\mathbb{R}[x_1,\ldots, x_n ]^G$.

\subsection{Settings for  quantum systems}

The mathematical structures associated with finite dimensional quantum systems,  in particularly with multi-qubit systems,  can be described as follows.

\noindent{ $\bullet$\, \bf  Quantum state space  $\bullet$ }
Introducing the space of $n\times n$ Hermitian matrices, $H_n\,,$ one can identify  the  density operators of an individual qubit  and  of a pair  of qubits
with a certain variety of   $H_2\, $ and $H_4\, $  respectively. In general, for $n\--$dimensional quantum system this variety, the \textit{state space}  $\mathfrak{P}(H_n),$  is given as the subset  of elements from $H_n\,,$  which  satisfy the  semipositivity and unit trace conditions:
\[
\label{eq:statespace}
\mathfrak{P}(H_n) := \{\varrho \in H_n  \mid   \varrho \geq 0\,, \mbox{tr}\varrho =1\}\,,
\]

\noindent{ $\bullet$\, \bf  Unitary symmetry of state space  $\bullet$ }
The traditional guiding philosophy to study physical models is based on the symmetry  principle.  In the case of quantum theory the basic symmetry is realized in the form of the adjoint action of the unitary group $U(n)$ on $H_n$:
\begin{equation}\label{eq:AdjTran}
(g, \varrho ) \to g\varrho g^\dag\, , \qquad  g \in U(n)\,, \quad \varrho \in H_n\,.
\end{equation}
Owing to this \textit{global  unitary symmetry}, the correspondence between states and physically relevant configurations is not one to one. All  density matrices, along the unitary  orbit,
\[
\mathcal{O}_\varrho = \{ g\varrho g^\dag, \forall  g \in SU(n) \}\,,
\]
represent one and  the same physical state.  The symmetry transformations (\ref{eq:AdjTran}) set the equivalence  relation
$\varrho \ {\sim} \  g\varrho g^\dag $ on  the state space  $\mathfrak{P}(H_n) \,,$
This equivalence defines the factor space  $\mathfrak{P}(H_n)/\sim$ and
allows    to ``reduce''  the  above outlined  ``redundant'' description of quantum system by passing to the   \textit{global unitary orbit space},  $\mathfrak{P}(H_n)/U(n)$\,.
The global unitary orbit space accumulates all physically relevant information  about the system as a whole.
Characteristics of  $\mathfrak{P}(H_n)/U(n)$ as an algebraic   variety   are encoded
in  the center of universal enveloping algebra $\mathfrak{U}(su(n))\,,$
and can be described in terms of the algebra of real $SU(n)-$invariant  polynomials in $\mathfrak{P}(H_n).$

\noindent{ $\bullet$\, \bf  Composite quantum systems $\bullet$ } If the space $H_n$ is associated with a composite quantum system, then another so-called \textit{ local unitary group} symmetry comes into play.
Restricting ourselves  to the case of  two-qubit system,  the local unitary group  is identified with the  subgroup $G=SU(2)\times SU(2) \subset SU(4)\,$ of the global unitary group $SU(4)\,.$
Opposite to the global unitary symmetry,  the local unitary group sets the equivalence  between states  of composite system which  have  one and the same entangling properties.   The algebra of corresponding
\textit{local unitary}  $G-$invariant polynomials
can be used for quantitative characterization of entanglement.
Having in mind application for the 2-qubit system
it is convenient to introduce  in this algebra of
local unitary $G\--$invariant polynomials  the $\mathbb{Z}^3-$grading.
This can be achieved by allocating  the algebra $\imath\mathfrak{su}(4)$ from $H_4$:
\[
\varrho=\frac{1}{4}\left[I_4+  \imath\mathfrak{su}(4)\right] \,,\qquad I_4\ \mbox{is identity $4\times 4$ matrix}
\]
and the decomposition of the latter into the direct sum of three real spaces
\begin{equation}\label{eq3decomp}
V_1= \imath\mathfrak{su}(2)\otimes I_2\,, \qquad V_2=I_2\otimes \imath\mathfrak{su}(2) \qquad
V_3= \imath\mathfrak{su}(2)\otimes \imath\mathfrak{su}(2)\,,
\end{equation}
 each representing a $G-$invariant subspace. Note that, if
the basis for the $\mathfrak{su}(2)$ algebra in each  subspace $V$ is chosen using the Pauli matrices
$\boldsymbol{\sigma}= (\sigma_1, \sigma_2, \sigma_3)\,,$ the  above
$G-$invariant  $\mathbb{Z}^3-$grading
gives
\begin{equation}\label{eq:Fanodecomp}
\varrho = \frac{1}{4}\left[
I_2\otimes I_2 +\sum_{i=1}^3 a_i \sigma_{i}\otimes{I}_2 + \sum_{i=1}^3b_i I_2 \otimes\sigma_{i} +
\sum_{i,j=1}^3 c_{ij}\sigma_i\otimes\sigma_j
\right]\,.
\end{equation}
This representation of 2-qubit state is
known as the Fano decomposition \cite{Fano}.
The real parameters $a_i, b_i$ and $ c_{ij}\,,  \, i,j=1,2,3\,,$
are subject to constraints coming from the semipositivity condition imposed
on the density matrix:
\begin{equation}\label{eq:semipos}
\varrho \geq 0\,.
\end{equation}
Explicitly, the semipositivity condition (\ref{eq:semipos})  reads  as a set of polynomial inequalities in the
fifteen  variables $a_i, b_i$ and $ c_{ij}\,$
(see, e.g., \cite{GerdtKhvedelidzePalii2009} and references therein).

\section{Applying invariant theory}

The entangling properties of composite quantum systems admit description within the general framework of the classical theory of invariants  (see books \cite{VinbergPopovItogi1989,DK} and references therein).

As it was mentioned above, for  the case of a 2-qubit the local unitary group is  $G=SU(2)\times SU(2)\,$.
The adjoint action   (\ref{eq:AdjTran}) of  this group on the 2-qubits density matrix $\rho$ induces  transformations on the space $W\,,$ defined
by the 15 real Fano variables (\ref{eq:Fanodecomp})\footnote{
More precisely, in correspondence with the  above mentioned $\mathbb{Z}^3-$ grading, the space $W$ is the representation space of irreducible representations of the form
$D_1\times D_0\,,$ $D_0\times D_1\,$ and $D_1\times D_1\,$  of   $SU(2)\times SU(2)$,
respectively.}
\begin{equation}\label{FanoParameters}
W:=\{\,(a_i,b_j,c_{kl})\in \R^{15} \mid i,j,k,l=1,2,3\,\}\,,
\end{equation}
and the corresponding  $G\--$invariant polynomials
accumulate all relevant information on the two-qubit entanglement.

Now we will give some known results on  the $G\--$invariant
polynomials ring structure. It is worth to note, that most  of these results are applicable for the linear actions of the compact groups on the linear spaces and thus cannot be directly  used for a description of quantum systems due to the  semipositivity of density matrix (\ref{eq:semipos}). However,  for a moment we relax the semipositivity constraints on  the Fano parameters and identify the space $W$  with $\mathbb{R}^{15}\,.$
The   positivity of density matrices can be written in the $G\--$invariant form and therefore can  be taken into account later.

\noindent{ $\bullet$\, \bf  The ring of $G-$invariant polynomials $\bullet$ }
Let $\mathbb{R} [W]:=\mathbb{R}[x_1, x_2, \dots, x_{15}]$ be  the coordinate ring of $W$ (\,with the ideal $I(W)=\{0\}$ in~\eqref{CoorRing}\,) and its
subring $R:={\R}[W]^G \subset \mathbb{R} [W]$ be the ring of polynomials invariant under the above mentioned  trans\-form\-ations on $W\,.$
  The invariant poly\-no\-mial ring $R$ has the following important properties~\cite{KingWelshJarvis,DK}.
\begin{itemize}
\item $R$ is a graded algebra over $\R$, and according to the  classical Hilbert theorem there is a finite set of homogeneous {\em fundamental invariants} generating ${R}$ as an $\R-$algebra.
\item The invariant ring $R$ is Cohen-Macaulay, that is, $R$ is finitely generated free module over $\R[F_p]$ (Hironaka decomposition)
    \[
       R=\bigoplus_{f_k\in F_s} f_k\, {\R}[F_p]\,.
    \]
    where $F_p$ is a set of algebraically independent {\em primary invariants} or {\em ho\-mo\-gene\-ous system of parameters}~\cite{DK} sometimes called {\em integrity basis} and $F_s$ is a set of linearly independent {\em secondary invariants}. In doing so, $1\in F_s$ and the set $F_p\cup F_s$ generates $R$.
\item Let $R_k$ be a subspace spanned by all homogeneous invariants in $R$ of degree $k$. If this subspace has dimension $d_k$, then the corresponding Molien series
    \begin{equation}\label{MFunction}
      M(q)=\sum_{k=0}^{\infty} d_k q^k
    \end{equation}
    generated by the Molien function $M(q)$ contains information on the number of primary and secondary invariants and their degrees (see formula (\ref{eq:Molien2qbit}) in the next Section).
\item Orbit separation:
   \[
   \forall u,v\in W\ \text{such that}\ G\cdot u \neq G\cdot v\ :\ \exists p\in R\
   \text{such that}\ p(u)\neq p(v)\,.
   \]
\end{itemize}
Because of the $G-$invariance of polynomials in $R$, their orbit separation property and Noetherianity of $R$, the use of fundamental invariants is natural in description of the orbit space of a linear action of a compact Lie group, and in particular of the $G-$invariant entanglement space of 2-qubit states.

\noindent{ $\bullet$\, \bf  Computational aspects $\bullet$} Constructive methods and algorithms for computing homogeneous generators of invariant rings are the main research objects of the computational invariant theory~\cite{DK,Sturmfels}. There are various algorithms known in the literature together with their im\-ple\-men\-ta\-tion in computer algebra software, e.g. {\sc Maple, Singular, Magma} (see book~\cite{DK}, Ch.3 and more recent paper~\cite{DK2008}). But, unfortunately, construction of a basis of invariants for $SU(2)\times SU(2)$ is too hard computationally for all those algorithms oriented to some rather wide classes of algebraic groups, and the integrity basis together with the secondary invariants  for the group has been constructed (see~\cite{KingWelshJarvis} and references therein) by the methods exploited its particular properties. We shall use this basis in the next sections.
Moreover, even our attempts to verify algebraic independence of the primary invariants, that is, to check that the variety in $\cC$ defined by polynomial set $F_p$ is $0$,  by using the standard \Gr basis technique for algebraic elimination, failed because of too high computer resources required.

\subsection{Basis of the ${SU(2)\times SU(2)}\--$invariant ring }

For two qubits the basis of  the polynomial ring   ${\R}[W]^{SU(2)\times SU(2)}$ was constructed in \cite{KingWelshJarvis}. The explicit form of its elements will be presented below.

As it was mentioned above, the space of polynomials in the fifteen variables (\ref{FanoParameters}) is decomposed into the irreducible representations of $SO(3) \times SO(3)$. Fur\-ther\-more, it inherits the  $\mathbb{Z}^3-$grading  in  $H_4$, and
 since the space of  homogeneous polynomials of degree $s,t,q$ in $a_i,b_i,c_{ij}\ (i,j=1,2,3)$, respectively, is invariant under the action of ${SU(2)\times SU(2)}$.  All such invariants $C$ can be classified according to their degrees $s,t,q$ of homogeneity in
$a_i,b_i,c_{ij}$.  Following Quesne's construction \cite{Quesne},  we shall denote them by ${C}^{(s\,t\,q)}$.
The degrees of homogeneous polynomials can be controlled from the knowledge of the Molien function.  The Molien function for mixed states of two qubits \cite{Quesne,GrasslRottelerBeth,KingWelshJarvis}:
\begin{equation}\label{eq:Molien2qbit}
M(q)=\frac{1+q^4+q^5+{3}q^6+{2}q^7+{2}q^8+
{3}q^9+q^{10}+q^{11}+q^{15}
}{(1-q)(1-q^2)^{3}(1-q^3)^{2}(1-q^4)^{3}(1-q^6)}\,,
\end{equation}
shows that integrity basis of the invariant ring
consists of 10 primary invariants of degrees
$1,2,2,2,3,3,4,4,4,6\,$, and there are 15 secondary invariants whose degrees are given by $
4,5,6,6,6,7,7,8,8,9,9,9,10,11,15\,$.
The Quesne invariants represent the resource of such primary and secondary invariants. Explicitly the Quesne  invariants read:
\begin{enumerate}
\item[\empty] 3 invariants of the second degree
\begin{eqnarray*}\label{eq:2order}
C^{(002)}=c_{ij}c_{ij}\,, \quad C^{(200)}=a_ia_i\,, \quad
C^{(020)}=b_ib_i\,,
\end{eqnarray*}
\item[\empty] 2 invariants of the third degree
\begin{eqnarray*}\label{eq:3order}
C^{(003)}= \frac{1}{3!}\epsilon_{ijk}\epsilon_{\alpha\beta\gamma}
c_{i\alpha}c_{j\beta}c_{k\gamma}\,, \qquad C^{(111)}=a_ic_{ij}b_j\,,
\end{eqnarray*}

\item[\empty] 4 invariants of the fourth degree
\begin{eqnarray*}\label{eq:4order}
  C^{(004)}&=&c_{i\alpha}c_{i\beta}c_{j\alpha}c_{j\beta}\,,\\
   C^{(202)}&=&a_i a_j c_{i\alpha}c_{j\alpha}\,,\\
   C^{(022)}&=&b_\alpha b_\beta c_{i\alpha}c_{i\beta}\,,\\
   C^{(112)}&=&\frac{1}{2}\,\epsilon_{ijk}\epsilon_{\alpha\beta\gamma}a_i b_\alpha
    c_{j\beta}c_{k\gamma}\,,
\end{eqnarray*}

\item[\empty] 1 invariant of the fifth degree
\begin{eqnarray*}\label{eq:5order}
&&C^{(113)}= a_i c_{i\alpha}c_{\beta \alpha}c_{\beta j}b_j\,,
\end{eqnarray*}

\item[\empty] 4  invariance of the six degree
\begin{eqnarray*}\label{eq:6order}
&&C^{(123)}=
    \epsilon_{ijk}b_i c_{\alpha j} a_\alpha c_{\beta k}c_{\beta l}b_l\,,\\
&&C^{(204)}=a_ic_{i\alpha}c_{j\alpha}c_{j\beta}c_{k\beta}a_k\,,\\
 &&C^{(024)}=b_ic_{\alpha i}c_{\alpha j}c_{\beta j}c_{\beta,
 k}b_k\,,\\
&&C^{(213)}=\epsilon_{\alpha\beta\gamma}a_\alpha c_{\beta i} b_i
c_{\gamma j} c_{\delta j} a_\delta\,,
\end{eqnarray*}
\item[\empty] 2  invariants of the seventh degree
\begin{eqnarray*}\label{eq:7order}
&&C^{(214)}=
    \epsilon_{ijk}b_i c_{\alpha j} a_\alpha c_{\beta k}c_{\beta l}c_{\gamma l}a_l\,,\\
&&C^{(124)}= \epsilon_{\alpha\beta\gamma}a_\alpha c_{\beta j} b_j
c_{\gamma k}c_{\delta k}c_{\delta l}b_l \,,
\end{eqnarray*}

\item[\empty] 2 invariants of the eights degree
\begin{eqnarray*}\label{eq:8order}
&&C^{(125)}=
    \epsilon_{ijk}b_i c_{\alpha j} c_{\alpha l}b_l  c_{\beta k}c_{\beta m}c_{\gamma m}
    a_\gamma\,,\\
&&C^{(215)}= \epsilon_{\alpha\beta\gamma}a_\alpha c_{\beta
i}c_{\delta i} a_\delta c_{\gamma k}c_{\varrho k}c_{\varrho l}b_l
\,,
\end{eqnarray*}

\item[\empty]2 invariants of the ninth degree
\begin{eqnarray*}\label{eq:9order}
&&C^{(306)}=
    \epsilon_{\alpha\beta\gamma}a_\alpha c_{\beta i}c_{\delta i}a_\delta
     c_{\gamma j}c_{\varrho j}c_{\varrho k}c_{\sigma k}a_\sigma\,,\\
&&C^{(036)}=\epsilon_{ijk}b_i c_{\alpha j}c_{\alpha l}b_l
     c_{\beta k}c_{\beta m}c_{\gamma m }c_{\gamma s}b_s\,,
\end{eqnarray*}
\end{enumerate}
In the above formulas the summation over all repeated indices from one to three is assumed.

\section{Constructing invariant polynomial ring of ${X-}$states}

Now we shall discuss  the  fate of the ${SU(2)\times SU(2)}\--$invariant polynomial ring, when the state space of two qubits is restricted to the
subspace of the $X\--$states.
We start with very brief settings of the $X-$states characteristics.

\subsection{$X-$states}

Consider subspace  $\mathfrak{P}_X\subset \mathfrak{P}(\mathbb{R}^{15})$,  of the
$X-$states. These states got such name due to the visual similarity of the density matrix, whose non-zero entries lie only on the main and minor (secondary) diagonals, with the Latin letter ``X'':
\begin{equation}
\label{eq:Xmatrix7}
\varrho_{X}:=
\left(
\begin{array}{cccc}
\varrho_{11}& 0 &0& \varrho_{14}\\
0&\varrho_{22} &\varrho_{23}& 0\\
0&\varrho_{32} &\varrho_{33}& 0\\
\varrho_{41}& 0 &0& \varrho_{44}
\end{array}
\right)\,.
\end{equation}
In (\ref{eq:Xmatrix7}) the diagonal entries
are real numbers,  while  elements of the minor diagonal are pairwise complex conjugated,  $\varrho_{14}=\overline{\varrho}_{14}$ and $\varrho_{23}=\overline{\varrho}_{32}\,.$

Comparing with the Fano decomposition (\ref{eq:Fanodecomp}) one can see, that
the $X\--$states belong to the 7-dimensional subspace $W_X$ of the vector space $W$
(\ref{FanoParameters}) defined as:
\[
 W_X := \{\,w \in W \ |\  c_{13}=c_{23}=c_{31}=c_{32}=0\,, a_i=b_i=0\,,\ i=1,2\,\}
\]
The $X-$matrices represent density operators that do not mix the subspaces cor\-re\-spond\-ing to matrix elements with indices 1-4 and 2-3 of the elements in the Hilbert space $\mathcal{H}_4$. It can be easily verified by using the permutation matrix
\[
P_{\pi}= \left[\begin{matrix}
1&0&0& 0\\
0&0&0&1\\
0&0&1&0\\
0&1&0&0
\end{matrix}
\right]
\]
that corresponds to the permutation
\[
\pi =\left(\begin{matrix}
1&2&3& 4\\
1&4&3&2
\end{matrix}\right )\,.
\]
The $X-$states can be transformed into the $2\times 2 $ block-diagonal form
\begin{equation}\label{eq:XmatrixBlockDiag}
\varrho_{X} =
P_{\pi}\left(
\begin{array}{cccc}
\varrho_{11}& \varrho_{14} &0& 0\\
\varrho_{41}& \varrho_{44}&0& 0\\
0& 0&\varrho_{33}&\varrho_{32} \\
0 & 0 &\varrho_{23}& \varrho_{22}\,.
\end{array}
\right)P_{\pi}\,.
\end{equation}

\subsection{Restriction of Quesne's invariants to the $X-$states subspace}

Now we consider the restriction of the above written fundamental invariants ${C}^{(s\,t\,q)}$ by Quesne to the subspace $W_X.$ The straightforward evaluation shows that the set of fundamental invariants
restricted to the $X-$state subspace $W_X\,,$
reduces to 12 nonzero invariants:
\begin{equation}\label{X-invariants}
{\small
\mathcal{P}=\{C^{200},C^{020},C^{002},
C^{111},C^{003},C^{202},
C^{022},C^{004},C^{112},
C^{113},C^{204},C^{024}\}.
}
\end{equation}
The explicit form of these invariants as polynomials in seven real variables, co\-or\-di\-nates on $W_X\,,$
\begin{equation}\label{X-parameters}
W_X:=\{\,(\alpha:=a_3,\beta:=b_3,\gamma:=c_{33},c_{11},c_{12},c_{21},c_{22})\in \R^7\,\}\,,
\end{equation}
is given by
\begin{eqnarray}
\mbox{deg}=2\, &&   C^{200}=\alpha^2, \quad
C^{020}=\beta^2, \quad
C^{002}=c_{11}^2+c_{12}^2+c_{21}^2+c_{22}^2+\gamma^2, \nonumber \\[0.2cm]
\mbox{deg}=3\, && C^{111}=\alpha\beta\gamma,\quad
C^{003}=\gamma(c_{11}c_{22}-c_{12}c_{21}), \nonumber \\[0.2cm]
\mbox{deg}=4\, &&
C^{202}=\alpha^2\gamma^2,\quad
C^{022}=\beta^2\gamma^2, \quad
C^{112}=\alpha\beta(c_{11}c_{22}-c_{12}c_{21}),\nonumber \\[0.2cm]
&& C^{004}=
\left(c_{11}^2+c_{12}^2+c_{21}^2+c_{22}^2 \right)^2
-2(c_{11}c_{22}-c_{12}c_{21})^2+\gamma^4,\nonumber \\[0.2cm]
\mbox{deg}=5\, && C^{113}=\alpha\beta\gamma^3,\nonumber \\[0.2cm]
\mbox{deg}=6\, && C^{204}=\alpha^2\gamma^4,\quad \nonumber
C^{024}=\beta^2\gamma^4.
\end{eqnarray}
Now, having the set of polynomials $\mathcal{P}$ in (\ref{X-invariants}), one can consider  the polynomial ring $\mathbb{R}[\mathcal{P}] \subset \mathbb{R}[W_X]$
generating by $\mathcal{P}$.\footnote{
Hereafter, slightly abusing notations we shall write $\R[W]$ and $\R[W_X]$ for the coordinate ring
of the  variety $W$ in~\eqref{FanoParameters} and its subvariety $W_X$ respectively.}

\subsection{Syzygy ideal in ${\mathbb{R}[\mathcal{P}]}$}

According to the  isomorphism (\ref{isomorphism}),  mentioned in the Section 2.1,  the  subring
$\mathbb{R}[\mathcal{P}_1, \dots ,  \mathcal{P}_{12}]$   can be written in the quotient form
\begin{equation}
\mathbb{R}[\mathcal{P}_1, \dots ,  \mathcal{P}_{12}]  \cong {\mathbb{R}[y_1, y_2\dots y_{12} ]}/{I_{\mathcal{P}}}\,, \label{InvRing}
\end{equation}
with the  syzygy ideal $I_{\mathcal{P}}$ for  $\mathcal{P}$
\[
I_{\mathcal{P}}: =\left\{\, h \in  \mathbb{R}[y_1, \dots,  y_{12}]\, \mid \, h( \mathcal{P}_1, \dots ,  \mathcal{P}_{12})\in  \mathbb{R}[W_X]\,\right\}\,.
\]

The syzygy ideal can be determined by applying the well-known elimination technique \cite{Sturmfels}.
Following this method we compute a \Gr basis of the ideal
\[
J_{\mathcal{P}}=\langle \mathcal{P}_1-y_1,\dots,  \mathcal{P}_{12}-y_{12} \rangle \subset \mathbb{R}[W_X,\mathcal{P}]
\]
for the lexicographic ordering
\begin{eqnarray*}
&& c_{11}\succ c_{12}\succ c_{21}\succ c_{22}\succ \alpha\succ \beta\succ \gamma\succ\\
&& \succ y_{12}\succ y_{11}\succ y_{10}\succ y_8\succ y_9\succ y_7\succ y_6\succ y_5\succ y_4\succ y_3\succ y_2\succ y_1\,.
\end{eqnarray*}
The intersection of the obtained \Gr basis with $\R[y_1,\ldots,y_{12}]$ forms a lexi\-cogra\-phic \Gr basis of the syzygy ideal $I_{\mathcal{P}}$. This basis consists of the following 37 polynomials
\begin{eqnarray*}
&I_{\mathcal{P}}=&\langle\, y_2y_6-y_4^2,\ y_1y_7-y_4^2\,,\ -y_1y_2y_5+y_4y_9\,,\ -y_1y_4y_5+y_6y_9\,,\\[0.1cm]
&& -y_2y_4y_5+y_7y_9\,,\ -y_1y_2y_3^2+y_1y_2y_8+2y_3y_4^2-2y_6y_7+2y_9^2\,, \\[0.1cm]
&& -y_1^2y_3^2y_4+y_1^2y_4y_8+2y_1^2y_5y_9+2y_1y_3y_4y_6-2y_4y_6^2\,, \\[0.1cm]
&& -y_2^2y_3^2y_4+y_2^2y_4y_8+2y_2^2y_5y_9+2y_2y_3y_4y_7-2y_4y_7^2\,, \\[0.1cm]
&& 2y_1^2y_2y_5^2-y_1y_3^2y_4^2+y_1y_4^2y_8+2y_3y_4^2y_6-2y_6^2y_7\,, \\[0.1cm]
&& 2y_1y_2^2y_5^2-y_2y_3^2y_4^2+y_2y_4^2y_8+2y_3y_4^2y_7-2y_6y_7^2\,, \\[0.1cm]
&& 2y_1y_2y_4^2y_5^2-y_3^2y_4^4+2y_3y_4^2y_6y_7+y_4^4y_8-2y_6^2y_7^2\,, \\[0.1cm]
&& 2y_1^3y_5^2-y_1^2y_3^2y_6+y_1^2y_6y_8+2y_1y_3y_6^2-2y_6^3\,, \\[0.1cm]
&&  2y_2^3y_5^2-y_2^2y_3^2y_7+y_2^2y_7y_8+2y_2y_3y_7^2-2y_7^3\,, \\[0.1cm]
&& y_1y_{10}-y_4y_6\,, y_{10}y_2-y_4y_7\,, y_{10}y_4-y_6y_7\,, \\[0.1cm]
&& y_1y_3^2y_4-y_1y_4y_8-2y_1y_5y_9-2y_3y_4y_6+2y_{10}y_6\,, \\[0.1cm]
&& y_2y_3^2y_4-y_2y_4y_8-2y_2y_5y_9-2y_3y_4y_7+2y_{10}y_7\,, -y_4^2y_5+y_{10}y_9\,, \\[0.1cm]
&& -2y_1y_2y_5^2+y_3^2y_4^2-2y_3y_6y_7-y_4^2y_8+2y_{10}^2\,, y_1y_{11}-y_6^2\,, y_{11}y_2-y_6y_7\,, \\[0.1cm]
&& y_1y_3^2y_4-y_1y_4y_8-2y_1y_5y_9-2y_3y_4y_6+2y_{11}y_4\,, \\[0.1cm]
&& -2y_1^2y_5^2+y_1y_3^2y_6-y_1y_6y_8-2y_3y_6^2+2y_{11}y_6\,, \\[0.1cm]
&& -2y_1y_2y_5^2+y_3^2y_4^2-2y_3y_6y_7-y_4^2y_8+2y_{11}y_7\,,\ -y_4y_5y_6+y_{11}y_9\,, \\[0.1cm]
&& y_1y_3^3y_4-y_1y_3y_4y_8-2y_1y_3y_5y_9-2y_1y_4y_5^2-y_3^2y_4y_6-y_4y_6y_8+2y_{10}y_{11}\,, \\[0.1cm]
&& -2y_1^2y_3y_5^2+y_1y_3^3y_6-y_1y_3y_6y_8-2y_1y_5^2y_6-y_3^2y_6^2-y_6^2y_8+2y_{11}^2\,, \\[0.1cm]
&& y_1y_{12}-y_6y_7,\, y_{12}y_2-y_7^2,\, y_2y_3^2y_4-y_2y_4y_8-2y_2y_5y_9-2y_3y_4y_7+2y_{12}y_4, \\[0.1cm]
&& -2y_1y_2y_5^2+y_3^2y_4^2-2y_3y_6y_7-y_4^2y_8+2y_{12}y_6\,, \\[0.1cm]
&& -2y_2^2y_5^2+y_2y_3^2y_7-y_2y_7y_8-2y_3y_7^2+2y_{12}y_7\,, -y_4y_5y_7+y_{12}y_9\,, \\[0.1cm]
&& y_2y_3^3y_4-y_2y_3y_4y_8-2y_2y_3y_5y_9-2y_2y_4y_5^2-y_3^2y_4y_7-y_4y_7y_8+2y_{10}y_{12}\,, \\[0.1cm]
&& -2y_1y_2y_3y_5^2+y_3^3y_4^2-y_3^2y_6y_7-y_3y_4^2y_8-2y_4^2y_5^2-y_6y_7y_8+2y_{11}y_{12}\,, \\[0.1cm]
&& -2y_2^2y_3y_5^2+y_2y_3^3y_7-y_2y_3y_7y_8-2y_2y_5^2y_7-y_3^2y_7^2-y_7^2y_8+2y_{12}^2\,\rangle.
\end{eqnarray*}
 Both {\sc Maple} and {\sc Mathematica} compute this basis in a few seconds on a PC. The ideal $I_{\mathcal{P}}$ has dimension 5. It is computed by the command {\tt HilbertDimension} in {\sc Maple}. To compute the dimension of $I_{\mathcal{P}}$  one can use the code available on the Mathematica Stack Exchange Web page {\tt http://mathematica.stackexchange.com/questions/37015/}.

If one uses the maximal independent set of variables $\{\,y_1,y_2,y_3,y_4,y_5\,\}$ as parameters, due to the above list of algebraic relations, the other variables are easily expressible in terms of the five parametric  variables by applying the command {\texttt{Solve}} in {\sc Maple} or {\sc Mathematica}:
\begin{eqnarray}
&& y_6 = \frac{y_4^2}{y_2}\,,\ \ y_7 = \frac{y_4^2}{y_1}\,, \nonumber \\[0.2cm]
&& y_8 = \frac{2y_1^3y_2^3y_5^2+y_1^2y_2^2y_3^2y_4^2-2y_1y_2y_3y_4^4+2y_4^6}{y_1^2y_2^2y_4^2}\,,\label{ResSyz}\\[0.2cm]
&& y_9 = \frac{y_1y_2y_5}{y_4}\,,\ \ y_{10} = \frac{y_4^3}{y_1y_2}\,,\ \ y_{11} = \frac{y_4^4}{y_1y_2^2}\,,\ \ y_{12} = \frac{ y_4^4}{y_1^2y_2}\,. \nonumber
\end{eqnarray}

This structure of the ring ${\mathbb{R}[y_1, y_2\dots y_{12} ]}/{I_{\mathcal{P}}}$
indicate  the fact that the polynomial invariants obtained from the rational relations \eqref{ResSyz}, by their conversion into  polynomials, form a \Gr basis of the syzygy ideal $I_{\mathcal{P}}$ in the ring of polynomials in $y_6,\ldots,y_{12}$ over the parametric coefficient field $\R(y_1,\ldots,y_5)$ of rational functions.

The determined properties of the ring $\mathbb{R}[\mathcal{P}_1, \dots ,  \mathcal{P}_{12}]$  is in a partial agreement with the initial structure of
 $\mathbb{R}[W]^{SU(2)\times SU(2)}$.  Indeed,  the  five
Quesne polynomials $C^{(200)}, C^{(020)}, C^{(002)}$, $C^{(111)}$ and $C^{(003)}$, that represent  the subset of the algebraically independent invariants, survive after restriction to the subspace $W_X$ and  correspond to the variables $y_1, y_2,  y_3, y_4, y_5$ which  are  independent modulo $I_{\mathcal{P}}$.
While restriction of the other Quesne invariants represent variables  which are  dependent modulo $I_{\mathcal{P}}$.


\subsection{Mapping ${\mathbb{R}[\mathcal{P}]}$ to a freely generating ring  }

Now  we establish the injective  homomorphism between  the ring $\mathbb{R}[\mathcal{P}]$ and a certain subring of  the coordinate ring $\mathbb{R}[W_X]$ which  is freely generated by polynomials  of degrees 1,1,1,2,2.
The latter subring is defined  as follows.
Consider the set of polynomials on $W_X$
\begin{equation}\label{eq:LIpol}
f_1=\gamma\,, \quad  g_1:=x_3+y_3, \quad g_2:=x_3-y_3,\  \quad g_3:=x_1^2+x_2^2\,,  \quad g_4:=y_1^2+y_2^2\,.
\end{equation}
where the following variables are introduced
\begin{eqnarray}
x_1 &=& c_{11}-c_{22}\,, \qquad  y_1  = c_{11}+c_{22}\,,\nonumber\\
x_2 &=&c_{12}+c_{21}\,, \qquad  y_2  = c_{12}-c_{21}\,,\label{eq:newcoordinatesW} \\
x_3&=&\alpha+\beta\,, \qquad\quad\  y_3  = \beta- \alpha \nonumber
\end{eqnarray}
It turns out that all twelve Quesne's polynomials $\mathcal{P}$ in (\ref{X-invariants})
can be expanded over these 5 algebraically independent polynomials.
The explicit form of these expansions for all non-vanishing Quesne's polynomials up to the order  six are given in the Table  1.

\noindent
Let us now introduce the ring $\mathbb{R}[f_1,  g_1, g_2, g_3, g_4]$, which is generated by the set  (\ref{eq:LIpol}).
The relation between the polynomials  of $\mathcal{P}$ in variables~\eqref{eq:LIpol} and Quesne's invariants, as shown in Table 1,  defines the mapping between the quotient ring of  $SU(2) \times SU(2)\--$invariant polynomials
and the ring $\mathbb{R}[f_1,  g_1, g_2, g_3, g_4]$
\begin{equation}\label{RingHOM}
 \phi:\ \R[y_1, y_2,  \dots, y_{12}]/I_{\mathcal{P}} \longrightarrow  \mathbb{R}[f_1,  g_1, g_2, g_3, g_4]\,,
\end{equation}
which is an injective ring homomorphism. Indeed, apparently mapping~\eqref{RingHOM} satisfies
\[
\phi(p+q)=\phi(p)+\phi(q)\,,\quad \phi(pq)=\phi(p)\phi(q)\,,
\]
and
\[
\phi(p)-\phi(q)=0\quad \text{if and only if}\quad p-q\in I_{\mathcal{P}}\,.
\]
However,~\eqref{RingHOM} is not isomorphism. The linear invariants $f,g_1,g_2$ have no preimages in $\mathbb{R}[\mathcal{P}]$ since  the polynomial invariants~\eqref{X-invariants} have degree $\geq 2$.

\begin{center}
\begin{table}[h]
\caption{Expansion of the Quesne's invariants for $X$-states.}
\begin{tabular}{|l||l||l||l|}
\hline
\hline
&&&\\
deg=2 &  $C^{200}=\displaystyle{\frac{1}{4}\,g_2^2}$ &
 $C^{020}=\displaystyle{\frac{1}{4}\,g_1^2}$ & $ C^{002}=\displaystyle{\frac{1}{2\,}(g_3+g_4)+f_1^2 }$\\
 &&&\\
\hline
\hline
 &&&\\
deg=3 & $ C^{111}=\displaystyle{\frac{1}{4}\,g_1g_2f_1}$ & $C^{003}=\displaystyle{\frac{1}{4}\,f_1(g_4-g_3)}$ &\\
 &&&\\
\hline
\hline
&&&\\
deg=4 & $ C^{202}=\displaystyle{\frac{1}{4}\,g_2^2f_1^2} $ & $ C^{022}=\displaystyle{\frac{1}{4}\,g_1^2f_1^2}  $ & $C^{004}=\displaystyle{
\frac{1}{8}\,(g_3+g_4)^2+\frac{1}{2}g_3g_4+f_1^4 } $ \\
&&&\\
&&&$ C^{112}=\displaystyle{\frac{1}{16}\,g_1g_2(g_4-g_3)}$\\
&&&\\
\hline
\hline
 &&&\\
deg=5& $C^{113}=\displaystyle{\frac{1}{4}\,g_1g_2f_1^3}$ &  &\\
 &&&\\
\hline
\hline
 &&&\\
deg=6 & $C^{204}=\displaystyle{\frac{1}{4}\,g_2^2f_1^4}$  & $C^{024}=\displaystyle{\frac{1}{4}\,g_1^2f_1^4}$ &\\
 &&&\\
\hline
\end{tabular}
\end{table}
\end{center}

\section{Concluding remarks}

We conclude with a group-theoretical explanation of the algebraic results
obtained in the previous section.
Note that the generic action of $SU(4)$ group on  subspace $\mathfrak{P}_X \subset \mathfrak{P}$  moves its elements from  $\mathfrak{P}_X\,.$
But one can point out the 7-dimensional subgroup $G_X \subset SU(4)$ that preserves the form of $X-$states.

\noindent{ $\bullet$\bf\, Invariance of  the $X-$states $\bullet$ }
The  7-parametric subgroup  $G_X \subset SU(4)$ that preserves $\mathfrak{P}_X$, i.e.,
\[
G_X \varrho_X G_X^\dag \in  \mathfrak{P}_X\,,
\]
can be easily constructed.  Let fix the  following  elements of the $SU(4)$ algebra
\footnote{The choice of such algebra generators is not unique, and there is the 15-fold degeneration: one can consider 15 different sets of seven generators that carry $X-$states into each other
\cite{Rau2009}.}
\begin{eqnarray*}
e_1 &=&\sigma_3\otimes \sigma_3,\ \\
e_2&=&\sigma_2\otimes \sigma_1, \
e_3=I \otimes \sigma_3,  \
e_4= - \sigma_2\otimes \sigma_2, \\
e_5&=&\sigma_1\otimes \sigma_2, \
e_6=\sigma_3\otimes I, \
e_7=\sigma_1\otimes \sigma_1 . \
\end{eqnarray*}
 The set of $( e_1, e_2, \dots, e_7)$ is closed under multiplication, i.e., it forms a basis of  the subalgebra
$\mathfrak{g}_X:= \mathfrak{su(2)}\oplus\mathfrak{u(1)}\oplus\mathfrak{su(2)}\in \mathfrak{su(4)}\,.$
Exponentiation   of  the algebra $\mathfrak{g}_X$ gives the  subgroup
\[
G_X :=\exp (i\mathfrak{g}_X)\in SU(4)\,,
\]
which is the invariance group of the $X-$states space  $\mathfrak{P}_X$.
Writing the generic element of algebra  $\mathfrak{g}_X$ as
$i \sum_i^7 \omega_ie_i\,,$
one can verify that an arbitrary element of  $G_X$  can be represented in the following
block-diagonal form
\begin{equation}\label{eq:G_XRepr}
G_X=P_{\pi}\left(
\begin{array}{c|c}
{e^{-i {\omega_1}}SU(2)}& 0  \\
\hline
0 &{e^{i {\omega_1}}SU(2)^\prime}\\
\end{array}
\right)P_{\pi}\,,
\end{equation}
where the two copies of $SU(2)$ are parametrized as follows
\begin{eqnarray*}
SU(2)&=& \exp{\left[i\left(\omega_4+\omega_7\right)\sigma_1 +i\left(\omega_2+\omega_5\right)\sigma_2 +
i\left(\omega_3+\omega_6\right)\sigma_3  \right]}\,,\\
SU(2)^\prime&=& \exp{\left[i\left(-\omega_4+\omega_7\right)\sigma_1 +i\left(-\omega_2+\omega_5\right)\sigma_2 +
i\left(\omega_3 -\omega_6\right)\sigma_3 \right] }\,.
\end{eqnarray*}

Having the representation (\ref{eq:G_XRepr}), one can find the transformation laws for
elements of  $X-$matrices.

\noindent{ $\bullet$\bf\,  Action $G_X$ on the $X-$states $\bullet$ }
First of all, the group $G_X$ leaves the parameter $c_{33}$ unchanged.
Secondly, according to  (\ref{eq:G_XRepr}), the adjoint action of the group $G_X$  induces the transformations of Fano parameters that    are unitary equivalent to the following block diagonal actions of two copies of $SO(3) $
on pair of 3-dimensional vectors in  $W$ with coordinates  (\ref{eq:newcoordinatesW}):
\[
\left(\begin{array}{c}
x_1^\prime  \\
x_2^\prime\\
x_3^\prime\\
y_1^\prime  \\
y_2^\prime\\
y_3^\prime\\
\end{array}
\right)=
\left(
\begin{array}{c|c}
&\\
\mbox{\Huge{SO(3)}}& \mbox{\Huge{O}}   \\
&\\
\hline
&\\
\mbox{\Huge{O}} &\mbox{\Huge{SO(3)}}^\prime\\
\end{array}
\right)
\left(
\begin{array}{c}
x_1  \\
x_2\\
x_3\\
y_1  \\
y_2\\
y_3
\end{array}\right)\,.
\]
Thus we conclude that there are 3  independent $G_X$-polynomial invariants
\[
f_1:= c_{33}\,,\quad f_2:=x_1^2+x_2^2+x_3^2\,,  \quad f_3:=y_1^2+y_2^2+y_3^2 \,.
\]
Similarly, the local transformation of the $X-$states can be identified  and the corresponding local unitary polynomial invariants can be determined.

\noindent{ $\bullet$\bf\,  Local subgroup of  $ G_X$  $\bullet$ }
One can easy verify that  the local subgroup of $ G_X$ is
\[
P_{\pi}\exp(\imath\ \frac{\varphi _1}{2}\sigma_3) \times \exp(\imath\frac{\varphi _2}{2} \sigma_3) P_{\pi} \subset G_X\,.
\]
Its action induces  two independent  $SO(2)$ rotations of two  planar vectors, \
$(x_1,\, x_2)$ and  $(y_1,\, y_2)$
on angles $\varphi_1+\varphi_2$ and $\varphi_1-\varphi_2\,$ respectively.
Therefore, the five polynomials (\ref{eq:LIpol}), used in the previous section for expansion of the $SU(2)\times SU(2)\--$invariants, represent
algebraically independent local invariants for the $X-$ states.

Concluding, our analysis of 2-qubits $X-$states space  shows the existence of two freely  generated polynomial rings, one related to the  global  $ G_X$-invariance
\[
 {\R}[W_X]^{G_X}={\R}[f_1, f_2, f_3]
\]
and another one, corresponding to the local unitary symmetry of $X\--$states,
\[
{\R}[W_X]^{SO(2)\times SO(2)}={\R}[f_1, g_1, g_2,  g_3, g_4]
\]
generated by the linear invariants $f_1,g_1,g_2$ together with the quadratic invariants $g_3,g_4$ of two planar vectors under the linear action of $SO(2)\times SO(2)$ group.

Moreover, the injective homomorphism of the ring of local unitary polynomial invariants,  ${\R}[W]^{SU(2)\times SU(2)}$ restricted to the subspace of 2-qubit  $X-$states,  to the above introduced freely generated
invariant ring ${\R}[W_X]^{SO(2)\times SO(2)}$ has been established.

\section*{Acknowledgments }
The authors thank S.Evlakhov for his help in some related computations. The contribution of the first author (V.G.) was partially supported by the Russian Foundation for Basic Research (grant 16-01-00080).


\noindent
{\small Vladimir P. Gerdt\\
Laboratory of Information Technologies,\\
Joint Institute for Nuclear Research, Dubna, Russia\\
Email: {\it gerdt@jinr.ru}
\vskip 0.2cm
\noindent
Arsen M. Khvedelidze\\
Laboratory of Information Technologies,\\
Joint Institute for Nuclear Research, Dubna, Russia\\
Institute of Quantum Physics and Engineering  Technologies,  \\
Georgian Technical University,\\ Tbilisi, Georgia\\
A Razmadze Mathematical Institute, \\
Iv. Javakhishvili, Tbilisi State University,\\
Tbilisi, Georgia\\
National Research Nuclear University,\\
 MEPhI (Moscow Engineering Physics Institute), Moscow, Russia\\
Email: {\it akhved@jinr.ru}
\vskip 0.2cm
\noindent
Yuri G. Palii\\
Laboratory of Information Technologies,\\
Joint Institute for Nuclear Research, Dubna, Russia\\
Institute of Applied Physics, \\
Chisinau, Republic of Moldova\\
Email: {\it palii@jinr.ru}
}

\end{document}